\documentclass[10pt,conference]{IEEEtran}

\usepackage{amssymb,amsfonts,amsmath,amstext,mathrsfs}
\usepackage{latexsym}
\usepackage{epsfig,psfrag,color,graphics,epsf}
\usepackage{enumerate}

\newtheorem{definition}{\textbf{Definition}}
\newtheorem{corollary}{\textbf{Corollary}}
\newtheorem{lemma}{\textbf{Lemma}}
\newtheorem{theorem}{\textbf{Theorem}}

\newtheorem{remark}{\textbf{Remark}}

\newcommand{\pp}{\hspace{1cm}}
\newcommand{\spp}{\hspace{5mm}}

\newcommand{\mE}{\mathrm{E}}
\newcommand{\Var}{\mathsf{Var}}

\newcommand{\cX}{\mathcal{X}}

\newcommand{\cY}{\mathcal{Y}}

\newcommand{\cW}{\mathcal{W}}

\newcommand{\cC}{\mathcal{C}}
\newcommand{\cN}{\mathcal{N}}

\newcommand{\cQ}{\mathcal{Q}}

\newcommand{\baralpha}{\bar{\alpha}}

\DeclareMathAlphabet{\matheuf}{U}{euf}{m}{n}
\newcommand{\eufC}{\mathscr{C}}

\IEEEoverridecommandlockouts

\begin{document}

\title{Secrecy Capacity Region of Binary
and Gaussian Multiple Access Channels
%
%
\thanks{This research was supported by the National Science Foundation
under Grant ANI-03-38807.} }
\author{\authorblockN{Yingbin Liang and H. Vincent Poor}
\authorblockA{Department of Electrical Engineering \\
Princeton University\\
Princeton, NJ 08544, USA \\
Email: \{yingbinl,poor\}@princeton.edu} }

\maketitle

\thispagestyle{plain}

\begin{abstract}
A generalized multiple access channel (GMAC) with one confidential
message set is studied, where two users (users 1 and 2) attempt to
transmit common information to a destination, and user 1 also has
confidential information intended for the destination. Moreover,
user 1 wishes to keep its confidential information as secret as
possible from user 2. A deterministic GMAC is first studied, and
the capacity-equivocation region and the secrecy capacity region
are obtained. Two main classes of the GMAC are then studied: the
binary GMAC and the Gaussian GMAC. For both channels, the
capacity-equivocation region and the secrecy capacity region are
established.
\end{abstract}

\section{Introduction}

An important security issue in multi-terminal networks is the
transmission of confidential information to legitimate
destinations while keeping other nodes as ignorant of this
information as possible. The secrecy level of a confidential
message at a nonlegitimate node (or a wire-tapper) is measured by
the equivocation rate, i.e., the entropy rate of the confidential
message conditioned on the channel outputs at this node. The
secrecy capacity is the maximum rate at which the confidential
message can be reliably transmitted to the intended destination
with the wire-tapper obtaining no information.

The secrecy capacity was established for a basic wire-tap channel
by Wyner in \cite{Wyner75}, and for a more general model of the
broadcast channel with confidential messages by
Csisz$\acute{\text{a}}$r and K$\ddot{\text{o}}$rner in
\cite{Csiszar78}. The relay channel with confidential messages was
studied in \cite{Oohama01}, where the secrecy rate was given. More
recently, a generalized multiple access channel (GMAC) with
confidential messages was studied in \cite{Liang06isit}, where
each user wishes to transmit a confidential message to a
destination, and wishes to keep the other user as ignorant of its
confidential message as possible. The secrecy rate region was
given for the GMAC with two confidential message sets, and the
secrecy capacity region was established for the GMAC with one
confidential message set. Other work on multiple access channels
with confidential messages can be found in \cite{Liu06,Tekin06}.

In this paper, we focus on the GMAC with one confidential message
set, where the two users have a common message for the destination
and only one user (user 1) has a confidential message for the
destination. We first study a simple deterministic GMAC, and
characterize the capacity-equivocation region and the secrecy
capacity region. The focus of this paper is on the two main
classes of GMACs: the binary GMAC and the Gaussian GMAC. For both
channels, we establish the capacity-equivocation region and the
secrecy capacity region explicitly.

In this paper, we use $x^n$ to indicate the vector
$(x_1,\ldots,x_n)$, and use $x_i^n$ to indicate the vector
$(x_i,\ldots,x_n)$. Throughout the paper, the logarithmic function
is to the base $2$.

The organization of this paper is as follows. In Section
\ref{sec:model}, we introduce the channel model of the GMAC with
one confidential message set. In Section \ref{sec:example}, we
present the secrecy capacity region of a simple deterministic GMAC
with one confidential message set. In Section \ref{sec:binary}, we
present the secrecy capacity region for a binary GMAC model with
one confidential message set. In Section \ref{sec:gauss}, we focus
on the Gaussian GMAC with one confidential message set, and
present our results on the secrecy capacity region.

\section{Channel Model and Previous Results}\label{sec:model}

In this section, we first define the GMAC with one confidential
message set, and then review the known results for this model.

\begin{figure}[tbhp]
\begin{center}
\includegraphics[width=8.5cm,clip]{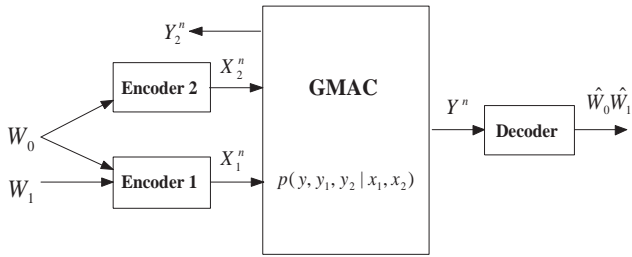}
\caption{GMAC with one confidential message set}
\label{fig:gmac1frame}
\end{center}
\end{figure}

\begin{definition}
A discrete memoryless GMAC consists of two finite channel input
alphabets $\cX_1$ and $\cX_2$, two finite channel output alphabets
$\cY$ and $\cY_2$, and a transition probability distribution
$p(y,y_2|x_1,x_2)$ (see Fig.~\ref{fig:gmac1frame}), where $x_1 \in
\cX_1$ and $ x_2 \in \cX_2$ are channel inputs from users 1 and 2,
respectively, and $y \in \cY$ and $ y_2 \in \cY_2$ are channel
outputs at the destination and user 2, respectively.
\end{definition}
\begin{definition}\label{def:phydegrade}
The GMAC with one confidential message set is physically degraded
if the transition probability distribution satisfies
\begin{equation}
p(y,y_2|x_1,x_2)= p(y|x_1,x_2)p(y_2|y,x_2),
\end{equation}
i.e., $y_2$ is independent of $x_1$ conditioned on $y$ and $x_2$.
\end{definition}
\begin{definition}
A $\left( 2^{nR_0},2^{nR_1}, n \right)$ code consists of the
following:
\begin{list}{$\bullet$}{\topsep=0ex \leftmargin=6mm
\rightmargin=2mm \itemsep=1mm}

\item Two message sets: $\cW_0=\{1,2,\ldots,2^{nR_0}\}$ and
$\cW_1=\{1,2,\ldots,2^{nR_1}\}$. The messages $W_0$ and $W_1$ are
independent and uniformly distributed over $\cW_0$ and $\cW_2$,
respectively.

\item Two (stochastic) encoders, one at user 1: $\cW_0 \times
\cW_1 \rightarrow \cX_1^n$, which maps each message pair
$(w_0,w_1) \in \cW_0 \times \cW_1$ to a codeword $x_1^n \in
\cX_1^n$; the other at user 2: $\cW_0 \rightarrow \cX_2^n$, which
maps each message $w_0 \in \cW_0$ to a codeword $x_2^n \in
\cX_2^n$;

\item One decoder at the destination: $\cY^n \rightarrow \cW_0
\times \cW_1$, which maps a received sequence $y^n$ to a message
pair $(w_0,w_1) \in \cW_0 \times \cW_1$.
\end{list}
\end{definition}

Note that although user 2 can receive channel outputs (see
Fig.~\ref{fig:gmac1frame}), it is only a passive listener in that
its encoding function is not affected by the received outputs.
However, since its outputs contain the confidential message $W_1$
sent by user 1, it may extract $W_1$ from its outputs. We assume
that user 1 treats user 2 as a wire-tapper, and wishes to keep it
as ignorant of $W_1$ as possible. The secrecy level of $W_1$ at
user 2 is measured by the following equivocation rate:
\begin{equation}
\frac{1}{n}H(W_1|Y_2^n,X_2^n,W_0).
\end{equation}
The larger the equivocation rate, the higher the level of secrecy.

A rate-equivocation triple $(R_0,R_1,R_{1,e})$ is {\em achievable}
if there exists a sequence of $\left(2^{nR_0}, 2^{nR_1}, n
\right)$ codes with the average error probability
$P_e^{(n)}\rightarrow 0$ as $n$ goes to infinity and with the
equivocation rate $R_{1,e}$ satisfying
\begin{equation}
R_{1,e} \leq \lim_{n\rightarrow \infty} \frac{1}{n}
H(W_1|Y_2^n,X_2^n,W_0).
\end{equation}
The rate-equivocation triple $(R_0,R_1,R_{1,e})$ indicates that
the rate pair $(R_0,R_1)$ can be achieved at the secrecy level
$R_{1,e}$ .

The capacity-equivocation region, denoted by $\eufC$, is the
closure of the set that consists of all achievable
rate-equivocation triples $(R_0,R_1,R_{1,e})$.

We are interested in the case where perfect secrecy is achieved,
i.e., user 2 does not get any information about the confidential
message that user 1 sends to the destination. This happens if
$R_e=R_1$.
\begin{definition}
The {\em secrecy capacity region} $\cC_s$ is the region that
includes all achievable rate pairs $(R_0,R_1)$ such that
$R_e=R_1$, i.e.,
\begin{equation}
\cC_s=\{(R_0,R_1): (R_0,R_1,R_1) \in \eufC \}.
\end{equation}
\end{definition}
\begin{definition}
For a given rate $R_0$, the {\em secrecy capacity} is the maximum
achievable rate $R_1$ with the confidential message perfectly
hidden from user 2, i.e.,
\begin{equation}
C_s (R_0)=\max_{(R_0,R_1) \in \cC_s } R_1 .
\end{equation}
\end{definition}

For the GMAC with one confidential message set, inner and outer
bounds on the capacity-equivocation region were given in
\cite{Liang06isit}. In particular, the exact secrecy capacity
region was established. For the degraded GMAC, the
capacity-equivocation region was established, which is given in
the following lemma. This lemma is useful to study binary and
Gaussian GMACs.
\begin{lemma}(\cite{Liang06isit})\label{lemma:dmcapa}
For the degraded GMAC with one confidential message set as in
Definition \ref{def:phydegrade}, the capacity-equivocation region
is given by
\begin{equation}\label{eq:dmcapa}
\begin{split}
& \eufC^d = \bigcup_{\begin{array}{l} p(q,x_2)p(x_1|q)
\\p(y|x_1,x_2) p(y_2|y,x_2)\end{array}} \\
& \left\{
\begin{array}{l}
(R_0,R_1,R_e): \\
R_0 \ge 0, R_1 \ge 0, \\
R_1 \leq I(X_1;Y |X_2,Q), \\
R_0+R_1 \leq I(X_1,X_2;Y), \\
0 \leq R_e \leq R_1, \\
R_e \leq I(X_1;Y |X_2,Q) - I(X_1;Y_2|X_2,Q), \\
R_0+R_e \leq I(X_1,X_2;Y)-I(X_1;Y_2 |X_2,Q) \\
\end{array} \right\}.
\end{split}
\end{equation}
where $Q$ is bounded in cardinality by $|\cQ |\leq |\cX_1|\cdot
|\cX_2|+1$.
\end{lemma}

\section{A Simple Example}\label{sec:example}

In this section, we consider a deterministic discrete memoryless
GMAC model with one confidential message set. We obtain the
capacity-equivocation region and the secrecy capacity region for
this channel.

\begin{figure}[t]
\begin{center}
\begin{psfrags}
\psfrag{x1x2}[c]{$X_1 \; X_2$}
\psfrag{y}[c]{$Y$}\psfrag{y2}[c]{$Y_2$}\psfrag{00}[c]{$0 \;\;\;
0$} \psfrag{01}[c]{$0\;\;\; 1$} \psfrag{10}[c]{$1\;\;\; 0$}
\psfrag{11}[c]{$1\;\;\; 1$} \psfrag{0}[c]{$0$} \psfrag{1}[c]{$1$}
\epsfig{file=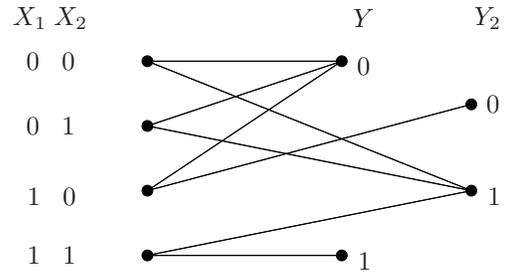,width=6cm}
\end{psfrags}
\caption{A deterministic GMAC model} \label{fig:bias}
\end{center}
\end{figure}

Consider a binary channel with all channel inputs and outputs
having alphabets $\{0,1\}$. The MAC from the two users to the
destination is a binary multiplier channel, and the channel from
user 1 to user 2 is a bias channel. The channel input-output
relationship (see Fig.~\ref{fig:bias}) is given by
\begin{equation}\label{eq:ex1}
Y=X_1 \cdot X_2, \pp Y_2=
\begin{cases}
1, \;\; & \text{if } X_1 \leq X_2; \\
0, & \text{if } X_1 > X_2.
\end{cases}
\end{equation}

\begin{figure}[t]
\begin{center}
\begin{psfrags}
\psfrag{r0}[c]{$R_0$} \psfrag{r1}[c]{$R_1$}\psfrag{1}[c]{$1$}
\epsfig{file=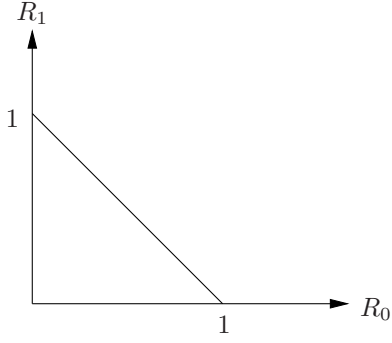,width=5cm}
\end{psfrags}
\caption{Secrecy capacity region of a deterministic GMAC }
\label{fig:multicapa}
\end{center}
\end{figure}

The capacity-equivocation region of the example channel given in
\eqref{eq:ex1} is:
\begin{equation}\label{eq:ex1capa}
\{ (R_0,R_1,R_e): R_0+R_1 \leq 1, R_e=R_1 \}.
\end{equation}
The capacity-equivocation region implies that the secrecy capacity
region of this channel is:
\begin{equation}\label{eq:ex1scr}
\{ (R_0,R_1): R_0+R_1 \leq 1 \}.
\end{equation}
Note that the region \eqref{eq:ex1scr} (see
Fig.~\ref{fig:multicapa}) coincides with the capacity region of
the binary multiplier MAC given in \cite{Meul73}.

To show that perfect secrecy can be achieved for all points in the
region \eqref{eq:ex1scr}, we first show that perfect secrecy can
be achieved for the two corner points. It is trivial that perfect
secrecy can be achieved for the corner point $(R_0=1,R_1=0)$,
i.e., $R_e=0$ is achievable at this point. For the other corner
point $(R_0=0,R_1=1)$, perfect secrecy is achieved by sending
$(x_1=0,x_2=1)$ for $W_1=0$ and $(x_1=1,x_2=1)$ for $W_1=1$. When
either of these two codewords is transmitted, user 2 always gets
output $Y_2=1$, and hence cannot determine whether $W_1=0$ or
$W_1=1$ is sent. Therefore, perfect secrecy is achieved. By
time-sharing between these two corner points, perfect secrecy can
be achieved for the entire region. Note that since the region
\eqref{eq:ex1capa} is the best possible rate-equivocation region
that can be achieved, it is hence the capacity-equivocation region
\eqref{eq:ex1capa}.
\begin{remark}
The deterministic GMAC defined in \eqref{eq:ex1} is a nondegraded
channel. We hence obtain the capacity-equivocation region for a
nondegraded channel.
\end{remark}

\section{The Binary GMAC with One Confidential Message
Set}\label{sec:binary}

In this section, we first follow \cite{Wyner73} to introduce
notation and useful lemmas for binary channels. We then introduce
the binary GMAC model we study and present the
capacity-equivocation region for this channel.

We first define the following operation:
\begin{equation}
a * b :=a(1-b)+(1-a)b \pp \text{ for } 0 \leq a, b \leq 1.
\end{equation}
We then define the following entropy function
\begin{equation}
h(a):=
\begin{cases}
-a\log a-(1-a)\log(1-a), \; & \text{if } 0 < a < 1; \\
0, & \text{if } a=0 \text{ or } 1.
\end{cases}
\end{equation}
Note that the function $h(a)$ is one-to-one for $0 \leq a \leq
1/2$. The inverse of the entropy function is limited to $h^{-1}(c)
\in [0,1/2]$.

\begin{lemma}\label{lemma:enconvex}(\cite{Wyner73})
The function $f(u)=h(\rho*h^{-1}(u)), 0 \leq u \leq 1$ (where
$\rho \in (0,1/2]$ is a fixed parameter) is strictly convex in
$u$.
\end{lemma}



The following useful lemma is a binary version of the entropy
power inequality.
\begin{lemma}\label{lemma:bepi}(\cite{Wyner73})
Consider two binary random vectors $X^n$ and $Y^n$. Let $H(X^n)\ge
nv$. Let
\begin{equation}
Y_i=X_i \oplus Z_i \pp \text{for } i=1,\ldots,n
\end{equation}
where $Z^n$ is a binary random vector with i.i.d.\ components and
$Z_i$ has distribution $Pr(Z_i=1)=p_0$ where $0 < p_0 \leq 1/2$.
The vectors $X^n$ and $Y^n$ can be viewed as inputs and outputs of
a binary symmetric channel (BSC) with the crossover probability
$p_0$. Then,
\begin{equation}
H(Y^n) \ge nh(p_0*h^{-1}(v))
\end{equation}
with equality if and only if $X^n$ has independent components, and
$H(X_i)=v$ for $i=1,2,\ldots,n$.
\end{lemma}

\begin{figure}[t]
\begin{center}
\begin{psfrags}
\psfrag{x1x2}[c]{$X_1 \; X_2$}
\psfrag{y}[c]{$Y$}\psfrag{y2}[c]{$Y_2$}\psfrag{00}[c]{$0 \;\;\;
0$} \psfrag{01}[c]{$0\;\;\; 1$} \psfrag{10}[c]{$1\;\;\; 0$}
\psfrag{11}[c]{$1\;\;\; 1$} \psfrag{0}[c]{$0$}
\psfrag{1p}[c]{$\scriptstyle 1-p$}\psfrag{p}[c]{$\scriptstyle
p$}\psfrag{1}[c]{$1$} \epsfig{file=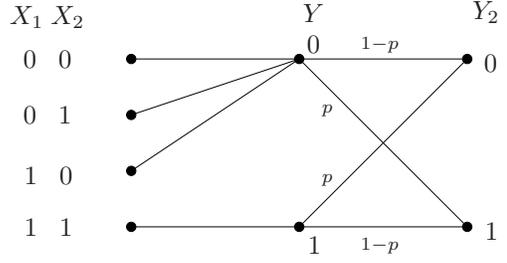,width=6cm}
\end{psfrags}
\caption{A degraded binary example GMAC} \label{fig:dmmac2}
\end{center}
\end{figure}

We now consider a discrete memoryless binary GMAC model with all
inputs and outputs having the binary alphabet set $\{0,1\}$. The
channel input-output relationship (see Fig.~\ref{fig:dmmac2}) at
each time instant satisfies
\begin{equation}\label{eq:bsc}
Y_i=X_{1,i} \cdot X_{2,i}, \spp Y_{2,i}= Y_i \oplus Z_{2,i} \spp
\text{for } i=1,\ldots,n
\end{equation}
where $Z_2^n$ is a binary random vector with i.i.d.\ components
and $Z_{2,i}$ has distribution $Pr(Z_{2,i}=1)=p$ where $0 < p \leq
1/2$. Note that the MAC channel from $(X_1,X_2)$ to $Y$ is a
binary multiplier channel. It is clear that this GMAC channel is
degraded, and the channel outputs $Y$ and $Y_2$ can be viewed as
the input and output of a discrete memoryless BSC with the
crossover probability $p$.

We have the following theorem on the capacity-equivocation region.
\begin{theorem}\label{theorem:bsccapa}
For the binary GMAC with one confidential message set defined in
\eqref{eq:bsc}, the capacity-equivocation region is
\begin{equation}\label{eq:bsccapa}
\eufC^B= \bigcup_{0 \leq \alpha \leq \frac{1}{2}} \left\{
\begin{array}{l}
(R_0,R_1,R_e): \\
R_0 \ge 0,R_1 \ge 0, \\
R_1 \leq h(\alpha), \\
R_0+R_1 \leq 1, \\
0 \leq R_e \leq R_1, \\
R_e \leq h(\alpha)+h(p)-h(p*\alpha), \\
R_0+R_e \leq 1+h(p)-h(p*\alpha)
\end{array} \right\}.
\end{equation}
\end{theorem}

The proof of Theorem \ref{theorem:bsccapa} is given at the end of
this section.

\begin{corollary}
The secrecy capacity region of the binary GMAC with one
confidential message set defined in \eqref{eq:bsc} is
\begin{equation}\label{eq:bsccrs}
\cC^B_s= \bigcup_{0 \leq \alpha \leq \frac{1}{2}} \left\{
\begin{array}{l}
(R_0,R_1): \\
R_0 \ge 0, R_1 \ge 0, \\
R_1 \leq h(\alpha)+h(p)-h(p*\alpha), \\
R_0+R_1 \leq 1+h(p)-h(p*\alpha)
\end{array} \right\}.
\end{equation}

The secrecy capacity as a function of $R_0$ is given by
\begin{equation}\label{eq:bsccrsr0}
C^B_s (R_0)= h(\alpha^*)+h(p)-h(p*\alpha^*)
\end{equation}
where $\alpha^*$ is determined by the following equation
\begin{equation}
R_0=1-h(\alpha^*).
\end{equation}
\end{corollary}

\begin{remark}
The BSC crossover probability parameter $p$ determines how noisy
the channel from user 1 to user 2 is compared to the channel from
user 1 to the destination. When $p=0$, user 2 has the same channel
from user 1 as the destination, and hence no secrecy can be
achieved. As $p$ increases, user 2 has a noisier channel from user
1 than the destination, and hence higher secrecy can be achieved.
As $p=\frac{1}{2}$, user 2 is totally confused by confidential
messages sent by user 1, and perfect secrecy is achieved.
\end{remark}
\begin{figure}[t]
\begin{center}
\begin{psfrags}
\epsfig{file=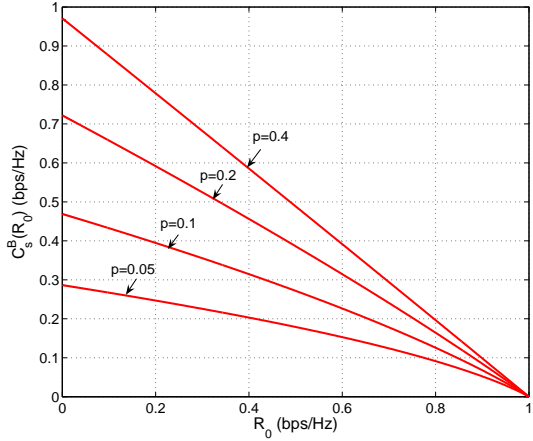,width=8cm}
\end{psfrags}
\caption{Secrecy capacity regions of the binary GMAC with one
confidential message set} \label{fig:bscr0e}
\end{center}
\end{figure}

Fig.~\ref{fig:bscr0e} plots the secrecy capacity as a function of
$R_0$ for four values of $p$. These lines of $C_s^B(R_0)$ also
serves as boundaries of the secrecy capacity regions with the
vertical axis being viewed as $R_1$. It is clear from
Fig.~\ref{fig:bscr0e} that as $p$ increases, the secrecy capacity
region enlarges, because user 2 is further confused about the
confidential message sent by user 1.

\begin{remark}\label{remark:twoschemes}
From the achievability proof of Theorem \ref{theorem:bsccapa}
(given at the end of this section), it can be seen that the
optimal scheme to achieve the secrecy capacity region uses
superposition encoding. To achieve the secrecy capacity
corresponding to different values of $R_0$, different values of
the superposition parameter $\alpha$ needs to be chosen to
generate the codebook. However, if the secrecy constraint is not
considered, the capacity region of the binary multiplier MAC can
be achieved by a time sharing scheme and superposition encoding is
not necessary.
\end{remark}
\begin{figure}[t]
\begin{center}
\begin{psfrags}
\epsfig{file=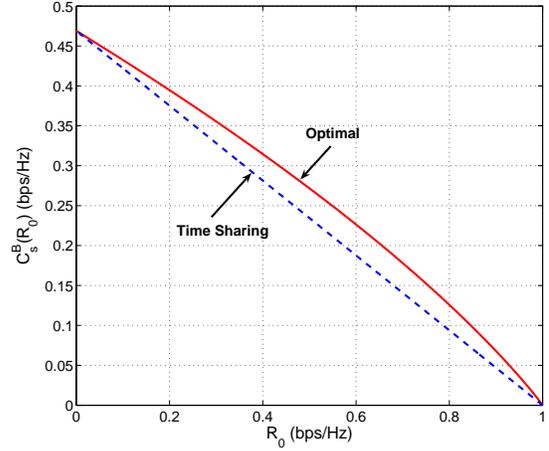,width=8cm}
\end{psfrags}
\caption{Comparison of secrecy capacity region and secrecy rate
region achieved by time sharing scheme for the binary GMAC with
one confidential message set} \label{fig:bsctwoschemes}
\end{center}
\end{figure}

Fig.~\ref{fig:bsctwoschemes} plots the secrecy capacity as a
function of $R_0$ (indicated by the solid line) and compares it
with the secrecy rate achieved by the time sharing scheme
(indicated by the dashed line). The figure demonstrates that the
time sharing scheme is strictly suboptimal to provide the secrecy
capacity region. As we commented in Remark
\ref{remark:twoschemes}, although the time sharing scheme is
optimal to achieve the capacity region of the binary multiplier
MAC, it is not optimal to achieve the secrecy capacity region of
the binary GMAC, where secrecy is also considered as a performance
criterion.

\vspace{0.2cm} \noindent{\em Proof of Theorem
\ref{theorem:bsccapa}}

\vspace{0.2cm} \noindent{\em Proof of Achievability}:

We apply Lemma \ref{lemma:dmcapa} to prove that the region
\eqref{eq:bsccapa} is achievable. Let $Q$ and $X'$ be two binary
random variables with alphabet $\{0,1\}$, and assume that $Q$ is
independent of $X'$. We choose the following joint distribution:
\begin{equation*}
\begin{split}
& Pr\{Q=0\}= \frac{1}{2}; \spp Pr\{X'=1\}=\alpha, \spp 0 \leq \alpha \leq \frac{1}{2}; \\
& Pr\{X_2=1 \} = 1; \spp  X_1=Q \oplus X'.
\end{split}
\end{equation*}

We now compute the mutual information terms in \eqref{eq:dmcapa}
given in Lemma \ref{lemma:dmcapa} based on the preceding joint
distribution.
\begin{equation*}
\begin{split}
R_1 & \leq I(X_1;Y |X_2,Q) =H(Y|X_2,Q) \\
& = Pr\{Q=0\}H(Y|X_2=1,Q=0)\\
& \spp +Pr\{Q=1\}H(Y|X_2=1,Q=1) \\
& = h(\alpha), \\
R_0 &+R_1 \leq I(X_1,X_2;Y) =H(Y)=1 ,\\
R_e & \leq I(X_1;Y |X_2,Q) - I(X_1;Y_2|X_2,Q) \\
& = h(\alpha)-(H(Y_2|X_2,Q)-H(Y_2|X_1,X_2)) \\
& = h(\alpha)-\big[Pr\{Q=0\}H(Y_2|X_2=1,Q=0)\\
& \hspace{1.3cm} +Pr\{Q=1\}H(Y_2|X_2=1,Q=1) \\
& \hspace{1.3cm} -Pr\{X_1=0\}H(Y_2|X_2=1,X_1=0)\\
& \hspace{1.3cm}-Pr\{X_1=1\}H(Y_2|X_2=1,X_1=1) \big]\\
& = h(\alpha)-[h(\alpha*p)-h(p)]\\
& = h(\alpha)+h(p)-h(\alpha*p),
\end{split}
\end{equation*}
\begin{equation*}
\begin{split}
R_0+R_1 & \leq I(X_1,X_2;Y)- I(X_1;Y_2|X_2,Q)\\
& =1-[h(\alpha*p)-h(p)] \\
& =1+h(p)-h(\alpha*p).
\end{split}
\end{equation*}

\noindent{\em Proof of the Converse:}


We consider a sequence of $\left( 2^{nR_0},2^{nR_1}, n \right)$
codes for the degraded GMAC with one confidential message set with
$P_e^{(n)} \rightarrow 0$. Then the probability distribution on
$W_0 \times W_1 \times \cX_1^n \times \cX_2^n \times \cY^n \times
\cY_2^n $ is given by
\begin{equation*}
\begin{split}
& p(w_0,w_1,x_1^n,x_2^n,y^n,y_2^n) \\
& \;=p(w_0)p(w_1)p(x_1^n |w_0,w_1)p(x_2^n|w_0) \prod_{i=1}^n
p(y_i,y_{2,i} | x_{1,i},x_{2,i})
\end{split}
\end{equation*}

From \cite[Sec.~4.2]{Liang06it}, we have the following bounds:
\begin{equation}\label{eq:conv1e}
\begin{split}
& nR_{1,e} \\
& \leq I(X_1^n;Y^n|X_2^n,W_0)-I(X_1^n;Y_2^n|X_2^n,W_0)+ n\delta_n \\
& \leq \sum_{i=1}^n
I(X_{1,i};Y_i|Q_i,X_{2,i})-I(X_{1,i};Y_{2,i}|Q_i,X_{2,i})+n\delta_n
\end{split}
\end{equation}
where $Q_i:=(Y^{i-1},X_2^n,W_0)$, and $\delta_n \rightarrow 0$ if
$P_e^{(n)} \rightarrow 0$.
\begin{equation}\label{eq:conv1e0}
\begin{split}
& nR_0+nR_{1,e} \\
& \leq I(W_0;Y^n)+
nR_{1,e}+n\delta_n \\
& = \sum_{i=1}^n
I(X_{1,i},X_{2,i};Y_i)-I(X_{1,i};Y_{2,i}|Q_i,X_{2,i})+n\delta_n
\end{split}
\end{equation}
\begin{equation}\label{eq:conv1}
\begin{split}
nR_1 & \leq I(X_1^n;Y^n|W_0,X_2^n) +n\delta_n \hspace{2.8cm}\\
& = \sum_{i=1}^n I(X_{1,i};Y_i |X_{2,i},Q_i)+ n\delta_n
\end{split}
\end{equation}
\begin{equation}\label{eq:conv10}
\begin{split}
nR_0+nR_1 & \leq I(W_0,W_1;Y^n) + n\delta_n \hspace{2.4cm} \\
& = \sum_{i=1}^n I(X_{1,i},X_{2,i};Y_i)+ n\delta_n.
\end{split}
\end{equation}

We now further derive the bounds
\eqref{eq:conv1e}-\eqref{eq:conv10} for the binary GMAC. From
\eqref{eq:conv1}, we obtain
\begin{equation}\label{eq:bscr1}
nR_1 \leq I(X_1^n;Y^n|X_2^n,W_0)+ n\delta_n =H(Y^n|X_2^n,W_0)+
n\delta_n
\end{equation}
where we have used the deterministic property of the GMAC, which
implies $H(Y^n|X_1^n,X_2^n,W_0)=0$.

Since $\{Y_i, 1 \leq i \leq n\}$ are binary random variables,
$H(Y_i) \leq 1$ for $1\leq i \leq n$. Hence
\begin{equation}
0 \leq H(Y^n|X_2^n,W_0) \leq \sum_{i=1}^n H(Y_i) \leq n.
\end{equation}
It is clear that there exists a parameter $\alpha \in [0,1/2]$
such that
\begin{equation}\label{eq:defalpha}
H(Y^n|X_2^n,W_0)= n h(\alpha).
\end{equation}
Substituting the preceding equation into \eqref{eq:bscr1}, we
obtain
\begin{equation}\label{eq:bconv1}
nR_1 \leq n h(\alpha)+ n\delta_n.
\end{equation}

From \eqref{eq:conv10}, we obtain
\begin{equation}\label{eq:bconv2}
\begin{split}
nR_0+nR_1 & \leq I(W_0,W_1;Y^n)+ n\delta_n \leq H(Y^n)+ n\delta_n
\\ & \leq n+ n\delta_n.
\end{split}
\end{equation}

From \eqref{eq:conv1e}, we obtain
\begin{equation}\label{eq:bscr1e}
\begin{split}
n&R_{1,e} \\
& \leq I(X_1^n;Y^n|X_2^n,W_0)-I(X_1^n;Y_2^n|X_2^n,W_0)+ n\delta_n \\
& = H(Y^n|X_2^n,W_0)-H(Y_2^n|X_2^n,W_0)\\
& \spp+H(Y_2^n|X_1^n,X_2^n,W_0) + n\delta_n \\
& \overset{(a)}{=} nh(\alpha)-H(Y_2^n|X_2^n,W_0) \\
& \spp + H(Y_2^n|Y^n,X_1^n,X_2^n,W_0)+ n\delta_n \\
& \overset{(b)}{=}  nh(\alpha)-H(Y_2^n|X_2^n,W_0)+ H(Y_2^n|Y^n)+ n\delta_n \\
& = nh(\alpha)-H(Y_2^n|X_2^n,W_0)+ nh(p) + n\delta_n \\
\end{split}
\end{equation}
In the preceding bound, the first term in $(a)$ follows from
\eqref{eq:defalpha}, the third term in $(a)$ follows from the fact
that $Y^n$ is a deterministic function of $(X_1^n,X_2^n)$, and the
third term in $(b)$ follows from the fact that $Y_2^n$ is
conditionally independent of everything else given $Y^n$.

Since $Z_2^n$ in \eqref{eq:bsc} is independent of $W_0,X_2^n$ and
$Y^n$, we apply Lemma \ref{lemma:bepi} to bound the term
$H(Y_2^n|X_2^n,W_0)$.
\begin{equation}\label{eq:bscy2}
\begin{split}
H &(Y_2^n|X_2^n,W_0) \\
& =\mE H(Y_2^n|X_2^n=x_2^n,W_0=w_0) \\
& \overset{(a)}{\ge} \mE
\left[nh\left(p*h^{-1}\left(\frac{H(Y^n|X_2^n=x_2^n,W_0=w_0)}{n}\right)
\right) \right]\\
& \overset{(b)}{\ge} nh\left(p*h^{-1}\left(\mE
\frac{H(Y^n|X_2^n=x_2^n,W_0=w_0)}{n}\right) \right) \\
& = nh\left(p*h^{-1}\left(
\frac{H(Y^n|X_2^n,W_0)}{n}\right) \right) \\
& \overset{(c)}{=} nh\left(p*h^{-1}\left(
\frac{nh(\alpha)}{n}\right) \right) \\
& = nh(p*\alpha)
\end{split}
\end{equation}
where $(a)$ follows from Lemma \ref{lemma:bepi}, $(b)$ follows
from Lemma \ref{lemma:enconvex} and Jensen's inequality, and $(c)$
follows from \eqref{eq:defalpha}.

Substituting \eqref{eq:bscy2} into \eqref{eq:bscr1e}, we obtain
\begin{equation}\label{eq:bconv3}
nR_{1,e} \leq nh(\alpha)+ nh(p)-nh(p*\alpha) + n\delta_n.
\end{equation}

From \eqref{eq:conv1e0}, we obtain
\begin{equation}\label{eq:bconv4}
\begin{split}
n&R_0+nR_{1,e} \\
& \leq I(W_0;Y^n)+nR_{1,e} \\
& \overset{(a)}{\leq} I(W_0;Y^n)+I(X_1^n;Y^n|X_2^n,W_0)\\
& \spp-I(X_1^n;Y_2^n|X_2^n,W_0)+ n\delta_n \\
& \overset{(b)}{\leq} I(W_0,X_2^n;Y^n)+I(X_1^n;Y^n|X_2^n,W_0)\\
&\spp -I(X_1^n;Y_2^n|X_2^n,W_0)+ n\delta_n \\
& = I(W_0,X_1^n,X_2^n;Y^n)-I(X_1^n;Y_2^n|X_2^n,W_0)+ n\delta_n \\
& = H(Y^n)-I(X_1^n;Y_2^n|X_2^n,W_0)+ n\delta_n \\
& \overset{(c)}{\leq} n+ nh(p)-nh(p*\alpha) + n\delta_n
\end{split}
\end{equation}
where $(a)$ follows from \eqref{eq:conv1e}, $(b)$ follows from the
chain rule and nonnegativity of mutual information, and $(c)$
follows from the steps in deriving $R_{1,e}$.

In summary, \eqref{eq:bconv1}, \eqref{eq:bconv2},
\eqref{eq:bconv3} and \eqref{eq:bconv4} constitute the converse
proof for Theorem \ref{theorem:bsccapa}.
\section{Gaussian GMAC with One Confidential
Message Set}\label{sec:gauss}

In this section, we study the Gaussian GMAC with one confidential
message set, where the channel outputs at the destination and user
2 are corrupted by additive Gaussian noise terms. We assume that
the channel is discrete and memoryless, and that the channel
input-output relationship at each time instant is given by
\begin{equation}\label{eq:gaussmodel}
\begin{split}
Y_i &=X_{1,i}+X_{2,i}+Z_i \\
Y_{2,i} &=X_{1,i}+X_{2,i}+Z_{2,i}
\end{split}
\end{equation}
where $Z^n$ and $Z_2^n$ are independent zero mean Gaussian random
vectors with i.i.d.\ components. We assume that $Z_i$ and
$Z_{2,i}$ have variances $N$ and $N_2$, respectively, where $N <
N_2$. The channel input sequences $X_1^n$ and $X_2^n$ are subject
to the average power constraints $P_1$ and $P_2$, respectively,
i.e.,
\begin{equation}\label{eq:powercons}
\frac{1}{n}\sum_{i=1}^n X_{1,i}^2 \leq P_1, \spp \text{and }\spp
\frac{1}{n}\sum_{i=1}^n X_{2,i}^2 \leq P_2.
\end{equation}

The following theorem states the capacity-equivocation region for
the Gaussian GMAC with one confidential message set.
\begin{theorem}\label{theorem:gausscapa}
For the Gaussian GMAC with one confidential message set given in
\eqref{eq:gaussmodel}, the capacity-equivocation region is given
by
\begin{equation}\label{eq:gausscapa}
\eufC^G= \bigcup_{0 \leq \alpha \leq 1} \left\{
\begin{array}{l}
(R_0,R_1,R_e): \\
R_0 \ge 0, R_1 \ge 0, \\
R_1 \leq \frac{1}{2}\log \left(1+\frac{\alpha P_1}{N}\right), \\
R_0+R_1 \leq \frac{1}{2}\log \left(1+\frac{P_1+P_2+2\sqrt{\baralpha P_1P_2}}{N}\right), \\
0 \leq R_e \leq R_1, \\
R_e \leq \frac{1}{2}\log \left(1+\frac{\alpha P_1}{N}\right)-\frac{1}{2}\log \left(1+\frac{\alpha P_1}{N_2}\right), \\
R_0+R_e \leq \frac{1}{2}\log
\left(1+\frac{P_1+P_2+2\sqrt{\baralpha P_1P_2}}{N}\right)\\
\hspace{1.8cm}-\frac{1}{2}\log \left(1+\frac{\alpha
P_1}{N_2}\right)
\end{array} \right\}.
\end{equation}
where $\baralpha=1-\alpha$ indicating the correlation between the
inputs from users 1 and 2.
\end{theorem}

The proof of Theorem \ref{theorem:gausscapa} is given at the end
of this section.
\begin{corollary}
The secrecy capacity region of the Gaussian GMAC with one
confidential message set given in \eqref{eq:gaussmodel} is
\begin{equation}\label{eq:gausscrs}
\cC^G_s= \bigcup_{0 \leq \alpha \leq 1} \left\{
\begin{array}{l}
(R_0,R_1): \\
R_0 \ge 0, R_1 \ge 0, \\
R_1 \leq \frac{1}{2}\log \left(1+\frac{\alpha P_1}{N}\right)-\frac{1}{2}\log \left(1+\frac{\alpha P_1}{N_2}\right), \\
R_0+R_1 \leq \frac{1}{2}\log
\left(1+\frac{P_1+P_2+2\sqrt{\baralpha P_1P_2}}{N}\right) \\
\hspace{1.8cm} -\frac{1}{2}\log \left(1+\frac{\alpha
P_1}{N_2}\right)
\end{array} \right\}.
\end{equation}

The secrecy capacity as a function of $R_0$ is
\begin{equation}\label{eq:gausscrsr0}
C^G_s (R_0)=
\begin{cases}
\frac{1}{2}\log \left(1+\frac{P_1}{N}\right)-\frac{1}{2}\log
\left(1+\frac{P_1}{N_2}\right), \\
\pp\pp \text{if } R_0 \leq \frac{1}{2}\log \frac{P_1+P_2+N}{P_1+N} \\
\frac{1}{2}\log \left(1+\frac{\alpha^*
P_1}{N}\right)-\frac{1}{2}\log \left(1+\frac{\alpha^*
P_1}{N_2}\right) \\
\pp\pp \text{if } R_0 > \frac{1}{2}\log \frac{P_1+P_2+N}{P_1+N}
\end{cases}
\end{equation}
where $\alpha^*$ is determined by the following equation
\begin{equation}
R_0 = \frac{1}{2}\log \frac{P_1+P_2+ 2\sqrt{(1-\alpha^*) P_1P_2}+
N}{\alpha^*P_1+N}.
\end{equation}
\end{corollary}

\begin{figure}[t]
\begin{center}
\begin{psfrags}
\epsfig{file=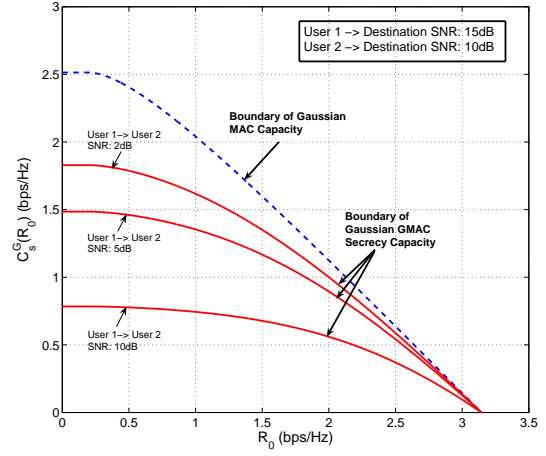,width=8cm}
\end{psfrags}
\caption{Secrecy capacity regions of Gaussian GMACs with one
confidential message set and capacity region of corresponding
Gaussian MAC} \label{fig:gaussr0e}
\end{center}
\end{figure}

Fig.~\ref{fig:gaussr0e} plots the secrecy capacity $C_s^G(R_0)$
(solid lines) of Gaussian GMACs with one confidential message set
for three user~1-to-user~2 SNR values. The lines of $C_s^G(R_0)$
also serve as boundaries of the secrecy capacity regions if we
view the vertical axis as $R_1$. It can be seen that as
user~1-to-user~2 SNR decreases, which implies that the noise level
at user 2 increases, user 2 gets more confused about the
confidential message sent by user 1. Thus the secrecy capacity
region enlarges. As this SNR approaches zero, the secrecy capacity
region approaches the entire capacity region of the Gaussian MAC,
which means that perfect secrecy is achieved for almost all points
in the capacity region of the MAC.

\vspace{0.2cm} \noindent{\em Proof of Theorem
\ref{theorem:gausscapa}}

We first note the following useful lemma.
\begin{lemma}\label{lemma:equcapa}(\cite{Liang06isit})
The capacity-equivocation region of GMACs with one confidential
message set depends only on the marginal channel transition
probability distributions $p(y|x_1,x_2)$ and $p(y_2|x_1,x_2)$.
\end{lemma}

To show Theorem \ref{theorem:gausscapa}, we first note that the
Gaussian GMAC defined in \eqref{eq:gaussmodel} is not physically
degraded according to Definition \ref{def:phydegrade}. However, it
has the same marginal distributions $p(y|x_1,x_2)$ and
$p(y_2|x_1,x_2)$ as the following physically degraded Gaussian
GMAC:
\begin{equation}\label{eq:dggaussmodel}
\begin{split}
Y_i &=X_{1,i}+X_{2,i}+Z_i \\
Y_{2,i} &=X_{1,i}+X_{2,i}+Z_i+Z'_i
\end{split}
\end{equation}
where $Z^n$ is the same as in \eqref{eq:gaussmodel}. The random
vector $Z'^n$ is independent of $Z^n$, and has i.i.d.\ components
with each component having the distribution $\cN(0,N_2-N)$.
According to Lemma \ref{lemma:equcapa}, it is sufficient to prove
Theorem \ref{theorem:gausscapa} for the physically degraded
Gaussian GMAC defined in \eqref{eq:dggaussmodel}.

\noindent{\em Proof of the Achievability:}

The achievability follows by computing the mutual information
terms in Lemma \ref{lemma:dmcapa} with the following joint
distribution:
\begin{equation}
\begin{split}
& Q =\phi, \pp X_2 \sim \cN(0,P_2) \\
& X'_1 \sim \cN(0,\alpha P_1), \text{ and $X'_1$ is independent of
$X_2$} \\
& X_1= \sqrt{\frac{\baralpha P_1}{P_2}}X_2+X'_1
\end{split}
\end{equation}

\noindent{\em Proof of the Converse:}

We apply the bounds \eqref{eq:conv1e}-\eqref{eq:conv10}, and
further derive these bounds for the degraded Gaussian GMAC.

From \eqref{eq:conv1}, we obtain
\begin{equation}
\begin{split}
nR_1 & \leq \sum_{i=1}^n I(X_{1,i};Y_i|X_{2,i},Q_i)+ n\delta_n \\
&  = \sum_{i=1}^n h(Y_i|X_{2,i},Q_i)-h(Y_i|X_{1,i},X_{2,i},Q_i)+
n\delta_n \\
& = \sum_{i=1}^n h(Y_i|X_{2,i},Q_i)-h(Z_i|X_{1,i},X_{2,i},Q_i)+
n\delta_n \\
& = \sum_{i=1}^n h(Y_i|X_{2,i},Q_i)-\frac{1}{2}\log 2\pi eN+
n\delta_n
\end{split}
\end{equation}

For the first term in the preceding inequality, we have
\begin{equation}\label{eq:y1}
\begin{split}
\sum_{i=1}^n & h (Y_i|X_{2,i},Q_i) \\
& = \sum_{i=1}^n h(X_{1,i}+X_{2,i}+Z_i|X_{2,i},Q_i) \\
& =\sum_{i=1}^n h(X_{1,i}+Z_i|X_{2,i},Q_i) \\
& \leq \sum_{i=1}^n h(X_{1,i}+Z_i) \leq \sum_{i=1}^n
\frac{1}{2} \log 2\pi e (\mE X_{1,i}^2+N) \\
& \overset{(a)}{\leq} \frac{n}{2} \log 2\pi e \left(\frac{1}{n}
\sum_{i=1}^n \mE X_{1,i}^2+N \right)\\
& \leq \frac{n}{2} \log 2\pi e (P_1+N)
\end{split}
\end{equation}
where $(a)$ follows from Jensen's inequality.

On the other hand,
\begin{equation}\label{eq:y2}
\begin{split}
\sum_{i=1}^n & h(Y_i|X_{2,i},Q_i) \\
& \ge \sum_{i=1}^n
h(X_{1,i}+X_{2,i}+Z_i|X_{1,i},X_{2,i},Q_i)\\
& = \frac{n}{2} \log 2\pi e N \; .
\end{split}
\end{equation}

Combining \eqref{eq:y1} and \eqref{eq:y2}, we establish that there
exists some $\alpha \in [0,1]$ such that
\begin{equation}\label{eq:alpha}
\sum_{i=1}^n h(Y_i|X_{2,i},Q_i) = \frac{n}{2} \log 2\pi e(\alpha
P_1+ N) \; .
\end{equation}

We hence obtain the bound for $R_1$
\begin{equation}
\begin{split}
nR_1 & \leq \frac{n}{2} \log 2\pi e(\alpha P_1+ N)-\frac{1}{2}\log
2\pi eN+ n\delta_n \\
& = \frac{n}{2} \log \left(1+\frac{\alpha P_1}{N}\right)+
n\delta_n \; .
\end{split}
\end{equation}

For the term $\sum_{i=1}^n h(Y_i|X_{2,i},Q_i)$, we can also derive
the following bound:
\begin{equation}
\begin{split}
& \sum_{i=1}^n h(Y_i|X_{2,i},Q_i) \\
& = \sum_{i=1}^n h(X_{1,i}+Z_i|X_{2,i},Q_i) \\
& \leq \sum_{i=1}^n \mE_{X_{2,i},Q_i} \frac{1}{2}\log 2\pi e
\Var(X_{1,i}+Z_i|X_{2,i},Q_i) \\
& \overset{(a)}{\leq } \sum_{i=1}^n \frac{1}{2}\log 2\pi e
\mE_{X_{2,i},Q_i} \Var(X_{1,i}+Z_i|X_{2,i},Q_i) \\
& = \sum_{i=1}^n \frac{1}{2}\log 2\pi e
\left(\mE_{X_{2,i},Q_i} \Var(X_{1,i}|X_{2,i},Q_i)+N \right)\\
& = \sum_{i=1}^n \frac{1}{2}\log 2\pi e
\left(\mE(X^2_{1,i})-\mE_{X_{2,i},Q_i}\mE^2(X_{1,i}|X_{2,i},Q_i) +N \right)\\
& \overset{(b)}{\leq } \frac{n}{2}\log 2\pi e
\Bigg( \frac{1}{n}\sum_{i=1}^n\mE(X^2_{1,i})\\
& \hspace{2.3cm} - \frac{1}{n}\mE_{X_{2,i},Q_i}\mE^2(X_{1,i}|X_{2,i},Q_i) +N \Bigg)\\
& \leq  \frac{n}{2}\log 2\pi e \left( P_1-
\frac{1}{n}\mE_{X_{2,i},Q_i}\mE^2(X_{1,i}|X_{2,i},Q_i) +N \right)
\end{split}
\end{equation}
where $(a)$ and $(b)$ follows from Jensen's inequality.

Using \eqref{eq:alpha}, we have
\begin{equation}\label{eq:eebound}
\begin{split}
& \alpha P_1+ N \leq P_1-
\frac{1}{n}\mE_{X_{2,i},Q_i}\mE^2(X_{1,i}|X_{2,i},Q_i) +N \\
\Longrightarrow \; &
\frac{1}{n}\mE_{X_{2,i},Q_i}\mE^2(X_{1,i}|X_{2,i},Q_i) \leq
\baralpha P_1
\end{split}
\end{equation}

From \eqref{eq:conv10}, we obtain
\begin{equation}
\begin{split}
nR_0+nR_1 & \leq \sum_{i=1}^n I(X_{1,i},X_{2,i};Y_i)+ n\delta_n \\
& =\sum_{i=1}^n h(Y_i)-I(Y_i|X_{1,i},X_{2,i})+ n\delta_n \\
& =\sum_{i=1}^n h(Y_i)-\frac{n}{2}\log 2\pi e N+ n\delta_n
\end{split}
\end{equation}

For the first term in the preceding inequality, we obtain
\begin{equation}
\begin{split}
& \sum_{i=1}^n h(Y_i) \\
& = \sum_{i=1}^n h(X_i+X_{1,i}+Z_i) \\
& \leq \sum_{i=1}^n \frac{1}{2} \log 2\pi e \left(
\mE(X_{1,i}+X_{2,i})^2+N \right) \\
& \overset{(a)}{\leq } \frac{n}{2} \log 2\pi e \left(
\frac{1}{n}\sum_{i=1}^n
\mE(X_{1,i}+X_{2,i})^2+N \right) \\
& \leq \frac{n}{2} \log 2\pi e \Bigg( \frac{1}{n}\sum_{i=1}^n \mE
X^2_{1,i} \\
& \hspace{2.1cm} + \frac{1}{n}\sum_{i=1}^n \mE X^2_{2,i}+
\frac{1}{n}\sum_{i=1}^n 2\mE (X_{1,i}X_{2,i}) +N \Bigg) \\
& \leq \frac{n}{2} \log 2\pi e \left( P_1+P_2+
\frac{1}{n}\sum_{i=1}^n 2\mE (X_{1,i}X_{2,i}) +N \right) \\
& \leq \frac{n}{2} \log 2\pi e \Bigg( P_1+P_2 \\
&\hspace{2.1cm}+
\frac{1}{n}\sum_{i=1}^n 2\mE\big(X_{2,i} \mE( X_{1,i}|X_{2,i},Q_i)\big) +N \Bigg) \\
& \overset{(b)}{\leq} \frac{n}{2} \log 2\pi e \Bigg( P_1+P_2 \\
& \hspace{2cm}+\frac{2}{n}\sum_{i=1}^n \sqrt{\mE X^2_{2,i} \cdot \mE \mE^2( X_{1,i}|X_{2,i},Q_i)} +N \Bigg) \\
& \overset{(c)}{\leq} \frac{n}{2} \log 2\pi e \Bigg( P_1+P_2 \\
& + 2 \sqrt{ \left(\frac{1}{n}\sum_{i=1}^n \mE X^2_{2,i} \right)
 \left(\frac{1}{n}\sum_{i=1}^n \mE \mE^2( X_{1,i}|X_{2,i},Q_i)\right)} +N \Bigg) \\
& \overset{(d)}{\leq} \frac{n}{2} \log 2\pi e \left( P_1+P_2+ 2
\sqrt{ \baralpha P_1 P_2} +N \right)
\end{split}
\end{equation}
In the preceding bound, $(a)$ follows from Jensen's inequality,
$(b)$ and $(c)$ follows from Cauchy-Schwarz inequality, and $(d)$
follows from \eqref{eq:eebound}.

Hence,
\begin{equation}
\begin{split}
nR_0&+nR_1 \\
& \leq \frac{n}{2} \log 2\pi e \left( P_1+P_2+ 2
\sqrt{
\baralpha P_1 P_2} +N \right)\\
& \spp-\frac{n}{2}\log 2\pi e N+ n\delta_n
\\
& = \frac{n}{2} \log \left(1+\frac{P_1+P_2+ 2 \sqrt{ \baralpha P_1
P_2}}{N} \right) + n\delta_n
\end{split}
\end{equation}

From \eqref{eq:conv1e}, we obtain
\begin{equation}\label{eq:re1a}
\begin{split}
& nR_{1,e} \\
& \leq \sum_{i=1}^n I(X_{1,i};Y_i|X_{2,i},Q_i)-I(X_{1,i};Y_{2,i}|X_{2,i},Q_i)+ n\delta_n \\
& =\frac{n}{2} \log \left(1+\frac{\alpha P_1}{N}\right)-\sum_{i=1}^n h(Y_{2,i}|X_{2,i},Q_i)\\
& \spp+ \frac{n}{2} \log 2\pi eN_2+ n\delta_n \\
\end{split}
\end{equation}

To bound the term $\sum_{i=1}^n h(Y_{2,i}|X_{2,i},Q_i)$ in
\eqref{eq:re1a}, we first derive the following bound. Since $Z'_i$
is independent of $Y_i$ given $X_{2,i}$ and $Q_i$, by entropy
power inequality, we obtain
\begin{equation*}
\begin{split}
& 2^{2 h(Y_i+Z'_i |X_{2,i}=x_{2,i},Q_i=q_i)}\\
& \;\; \ge 2^{2 h(Y_i |X_{2,i}=x_{2,i},Q_i=q_i)}+ 2^{2 h(Z'_i
|X_{2,i}=x_{2,i},Q_i=q_i)} \\
& \;\;= 2^{2 h(Y_i |X_{2,i}=x_{2,i},Q_i=q_i)}+ 2\pi e (N_2-N)
\end{split}
\end{equation*}
We then obtain
\begin{equation*}
\begin{split}
& h(Y_i+Z'_i |X_{2,i}=x_{2,i},Q_i=q_i) \\
&\;\; \ge \frac{1}{2} \log \left(2^{2 h(Y_i
|X_{2,i}=x_{2,i},Q_i=q_i)} + 2\pi e (N_2-N) \right)
\end{split}
\end{equation*}
Taking the expectation on both sides of the preceding equation, we
obtain
\begin{equation*}
\begin{split}
\mE & h(Y_i+Z'_i |X_{2,i}=x_{2,i},Q_i=q_i) \\
& \ge \frac{1}{2} \mE
\log \left(2^{2 h(Y_i |X_{2,i}=x_{2,i},Q_i=q_i)} + 2\pi e
(N_2-N) \right) \\
& \overset{(a)}{\ge} \frac{1}{2} \log \left(2^{2 \mE h(Y_i
|X_{2,i}=x_{2,i},Q_i=q_i)} + 2\pi e
(N_2-N) \right) \\
& = \frac{1}{2} \log \left(2^{2 h(Y_i |X_{2,i},Q_i)} + 2\pi e
(N_2-N) \right)
\end{split}
\end{equation*}
where $(a)$ follows from Jensen's inequality and the fact that
$\log(2^x+c)$ is a convex function.

Summing over the index $i$, the preceding inequality becomes
\begin{equation*}
\begin{split}
\sum_{i=1}^n & h(Y_i+Z'_i |X_{2,i},Q_i) \\
& \ge
\frac{1}{2}\sum_{i=1}^n \log \left(2^{2 h(Y_i |X_{2,i},Q_i)} +
2\pi e (N_2-N) \right) \\
& \overset{(a)}{\ge} \frac{n}{2} \log \left(2^{2
\frac{1}{n}\sum_{i=1}^n h(Y_i
|X_{2,i},Q_i)} + 2\pi e (N_2-N) \right) \\
& \overset{(b)}{=} \frac{n}{2} \log \left( 2\pi e (\alpha P_1+ N)+ 2\pi e (N_2-N) \right) \\
& = \frac{n}{2} \log \left( 2\pi e (\alpha P_1+ N_2) \right) \\
\end{split}
\end{equation*}
where $(a)$ follows from Jensen's inequality, and $(b)$ follows
from \eqref{eq:eebound}.

Applying the preceding bound to the term $\sum_{i=1}^n
h(Y_{2,i}|X_{2,i},Q_i)$, we obtain
\begin{equation}
\begin{split}
\sum_{i=1}^n h(Y_{2,i}|X_{2,i},Q_i)& =\sum_{i=1}^n
h(Y_i+Z'_i|X_{2,i},Q_i) \\
& \ge \frac{n}{2} \log \left( 2\pi e (\alpha P_1+ N_2) \right)
\end{split}
\end{equation}

Substituting the preceding bound into \eqref{eq:re1a}, we obtain
\begin{equation}
\begin{split}
nR_{1,e} & \leq \frac{n}{2} \log \left(1+\frac{\alpha P_1}{N}
\right) \\
& \spp-\frac{n}{2} \log \left( 2\pi e (\alpha P_1+ N_2) \right)+ \frac{n}{2} \log 2\pi eN_2+ n\delta_n \\
& \leq  \frac{n}{2} \log \left(1+\frac{\alpha P_1}{N} \right)
-\frac{n}{2} \log \left( 1+ \frac{\alpha P_1}{ N_2}\right) + n\delta_n \\
\end{split}
\end{equation}

From \eqref{eq:conv1e0}, we obtain
\begin{equation}
\begin{split}
n & R_0 +nR_{1,e} \\
& \leq \sum_{i=1}^n I(X_{1,i},X_{2,i};Y_i)-I(X_{1,i};Y_{2,i}|X_{2,i},Q_i)+ n\delta_n \\
& \leq \frac{n}{2} \log \left(1+\frac{P_1+P_2+ 2 \sqrt{ \baralpha
P_1 P_2}}{N} \right)\\
& \spp-\sum_{i=1}^n h(Y_{2,i}|X_{2,i},Q_i)+ \frac{n}{2}
\log 2\pi eN_2+ n\delta_n  \\
& \leq \frac{n}{2} \log \left(1+\frac{P_1+P_2+ 2 \sqrt{ \baralpha
P_1 P_2}}{N} \right)\\
& \spp -\frac{n}{2} \log \left( 2\pi e (\alpha P_1+ N_2) \right)+
\frac{n}{2}
\log 2\pi eN_2+ n\delta_n  \\
& \leq \frac{n}{2} \log \left(1+\frac{P_1+P_2+ 2 \sqrt{ \baralpha
P_1 P_2}}{N} \right)\\
& \spp -\frac{n}{2} \log ( 1+ \frac{\alpha P_1}{ N_2})+ n\delta_n,
\end{split}
\end{equation}
which completes the proof.

\section{Conclusions}

We have established the capacity-equivocation region for a binary
example GMAC and the Gaussian GMAC with one confidential message
set. For the binary GMAC, we have shown that the time-sharing
scheme is strictly suboptimal to achieve the secrecy capacity,
although it is optimal to achieve the capacity without the secrecy
constraint. We have also found that the capacity-equivocation
region of GMACs with one confidential message set depends only on
the marginal channels $p(y|x_1,x_2)$ and $p(y_2|x_1,x_2)$. Based
on this observation, we have obtained the capacity-equivocation
region for the Gaussian GMAC (not necessarily physically degraded)
with one confidential message set.

\bibliographystyle{IEEEtran}

\end{document}